\documentclass[prl,twocolumn,showpacs,preprintnumbers,amsmath,amssymb]{revtex4}
\usepackage{graphicx}

\begin{document}
\title{Superconducting $\pi$ qubit with a ferromagnetic Josephson junction}
\author{T. Yamashita,$^{1}$ K. Tanikawa,$^{1}$ S. Takahashi,$^{1}$ and S. Maekawa$^{1,2}$}
\affiliation{$^{1}$Institute for Materials Research, Tohoku University, 
Sendai, Miyagi, 980-8577, Japan \\
$^{2}$CREST, Japan Science and Technology Agency (JST), 
Kawaguchi, Saitama, 332-0012, Japan}
\pacs{03.67.Lx, 74.50.+r, 74.45.+c, 85.25.Cp}

\begin{abstract}
Solid-state qubits have the potential for the large-scale integration 
and for the flexibility of layout for quantum computing.  
However, their short decoherence time due to the coupling to the environment 
remains an important problem to be overcome.  
We propose a new superconducting qubit which incorporates a spin-electronic device: 
the qubit consists of a superconducting ring 
with a ferromagnetic $\pi$ junction which has a metallic contact and 
a normal Josephson junction with an insulating barrier.  
Thus, a quantum coherent two-level state is formed without an external magnetic field.  
This feature and the simple structure of the qubit make it possible to reduce its size 
leading to a long decoherence time.  
\end{abstract}

\maketitle

The quantum computer is an innovative device in the sense that it would make it 
possible to solve problems which require unrealistically long computation times 
on a classical computer \cite{Nielsen}.  
In the quantum computer, the information is stored 
in a basic element called the qubit, which is a quantum coherent two-level system.  
The superposition of the two-level state is utilized in the process of quantum computing.  
For the physical realization of the qubit, 
various systems have been proposed, e.g., ion traps, nuclear spins, and photons.  
Among the proposals, solid-state devices have the advantage of large-scale
integration and flexibility of layout. On the other hand, a challenging problem for the
solid state qubits is the reduction of the decoherence effect, since the solid states qubits in
general have a short decoherence time due to their coupling to the environment.
In recent years, several qubits based on the Josephson effect have been proposed 
\cite{Nakamura,Vion,Pashkin,Yamamoto,Yu,Mooij,Orlando,Wal,Chiorescu,Ioffe,Blatter,Makhlin}.  
One of the proposals involves a Cooper-pair box type of qubit 
\cite{Nakamura,Vion,Pashkin,Yamamoto}.  
In this case, quantum oscillations between the quantum two-level states (Rabi oscillations) 
have been detected \cite{Nakamura,Vion}, 
and the operation of coupled two qubits has been demonstrated \cite{Pashkin,Yamamoto}.  
Another example is a flux qubit which uses the superconducting phase.  
For this proposal, a circuit with a single and 
relatively large Josephson junction has been demonstrated \cite{Yu}.  
Mooij {\it et al.} have also proposed a flux qubit which consists of 
a superconducting loop with three Josephson junctions \cite{Mooij,Orlando,Wal,Chiorescu}. 
In this qubit, degenerate double minima form in the superconducting phase space, 
when an external magnetic field, which corresponds to the half of the unit magnetic flux, 
is applied.  
The bonding and antibonding states which form due to the tunneling between these two states 
can be used as the two quantum coherent states.  
Experimentally, microwave-induced transitions between the two quantum states 
have been observed for this qubit \cite{Mooij,Orlando,Wal,Chiorescu}.  
Another proposal is a qubit which does not require an external magnetic field and 
uses an {\it s}-wave/{\it d}-wave/{\it s}-wave ({\it sds}) superconducting junction 
\cite{Ioffe}.  
In addition, a five-junction device 
with one ferromagnetic $\pi$ junction and four normal Josephson junctions 
has been discussed in analogy with the ${\it sds}$ qubit \cite{Blatter}.  

Recent advances in microprocessing techniques have yielded a variety of 
spin-electronic devices \cite{Maekawa}.  
For instance, in ferromagnet/superconductor (FM/SC) junctions, novel phenomena 
due to the competition between ferromagnetism and superconductivity are expected 
\cite{Soulen,Strikers}.  
There have also been extensive theoretical studies of Josephson $\pi$ junctions 
consisting of a ferromagnet sandwiched between two superconductors (SC/FM/SC) 
\cite{Buzdin,Bulaevskii,Golubov,Radovic,Radovic2,Demler}.  
In this respect, several experimental observations of the $\pi$ state have been reported 
\cite{Ryazanov1,Kontos0,Kontos1,Kontos2,Ryazanov2,Bauer}.  
At the interface between an SC and an FM, 
Cooper pairs penetrating into the FM have a finite momentum $Q \propto h_{ex}/v_{F}$, 
where $v_{F}$ is the Fermi velocity, because of the exchange splitting $h_{ex}$ 
between the up and the down spin bands. \cite{Demler}.  
Consequently, the superconducting order parameter $\psi$ oscillates as 
$\psi \propto \cos{(2Qx)}$ along the direction $x$ which is perpendicular to the interface.  
In SC/FM/SC junctions, the order parameters in the two SC's take different signs, 
when the thickness of the FM is about half of the period of the oscillation.  
This causes the current-phase relation to be shifted by 
$\pi$ from that of a normal Josephson junction.  
This is the so-called $\pi$ junction \cite{Buzdin,Bulaevskii,Golubov,Radovic,Radovic2,Demler,
Ryazanov1,Kontos0,Kontos1,Kontos2,Ryazanov2,Bauer}.  

\begin{figure}[t]
\centering
\includegraphics[width=\columnwidth]{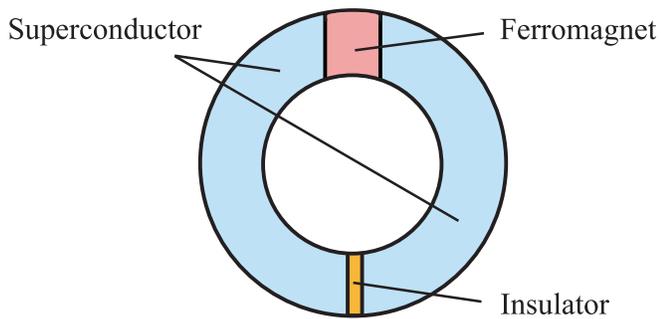}
\caption{Schematic diagram of a superconducting ring with an insulator and a ferromagnet.  
The superconductor/insulator/superconductor (SC/I/SC) junction is a $0$ junction 
(a normal Josephson junction), and the superconductor/ferromagnet/superconductor (SC/FM/SC) 
junction is a metallic $\pi$ junction.}
\label{Fig_G}
\end{figure}
In this Letter, we study a qubit which consists of a superconducting ring 
with an insulator and a ferromagnet as shown in Fig. \ref{Fig_G}.  
In the ring, the superconductor/insulator/superconductor (SC/I/SC) junction is 
a normal Josephson junction with Josephson energy $U_0 = - E_0 \cos{\theta}$.  
Here, $E_0$ is the coupling constant in the SC/I/SC junction and 
$\theta$ is the phase difference between the SC's.  
On the other hand, the SC/FM/SC junction is a metallic $\pi$ junction.  
Before starting the discussion on the qubit, 
it is useful to first derive the Josephson energy $U_{\pi}$ of the metallic $\pi$ junction, 
by using the Bogoliubov-de Gennes (BdG) equation \cite{BdG}.  
Solving the BdG equation, 
we obtain the Andreev bound state energy $E_{\sigma}$ for spin $\sigma$ \cite{Furusaki}.  
The phase, $\theta$, dependent part of the free energy in the SC/FM/SC junction 
is then expressed in terms of $E_{\sigma}$ as 
\begin{eqnarray}
F 
&=& - k_B T\sum\limits_\sigma {\sum\limits_{E_\sigma   > 0} 
{\ln \left[ {2\cosh \left( {E_\sigma  /2k_B T} \right)} \right]} } .  
\label{FE1}
\end{eqnarray}
As usual, $T$ is the temperature and $k_B$ is the Boltzmann constant.  
At low temperatures, the free energy (Eq. (\ref{FE1})) reduces to 
\begin{equation}
F = - \frac{1}{2}\sum\limits_\sigma {\sum\limits_{E_\sigma > 0} {E_\sigma}}.  
\label{FE2}
\end{equation}

\begin{figure}[t]
\centering
{\includegraphics[width=0.85\columnwidth]{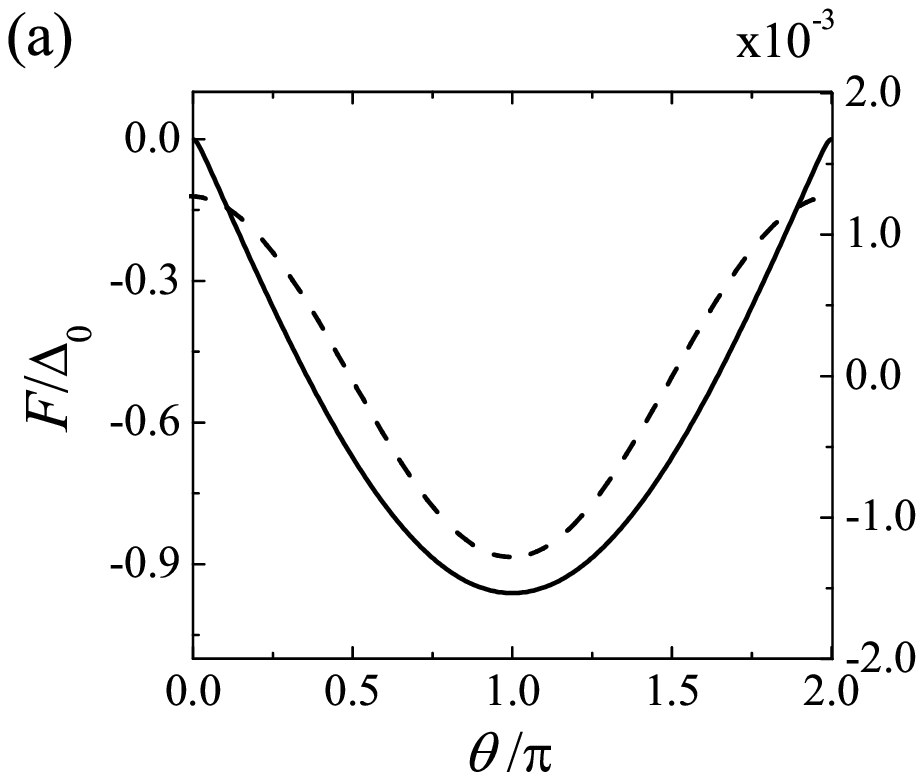}
\includegraphics[width=0.85\columnwidth]{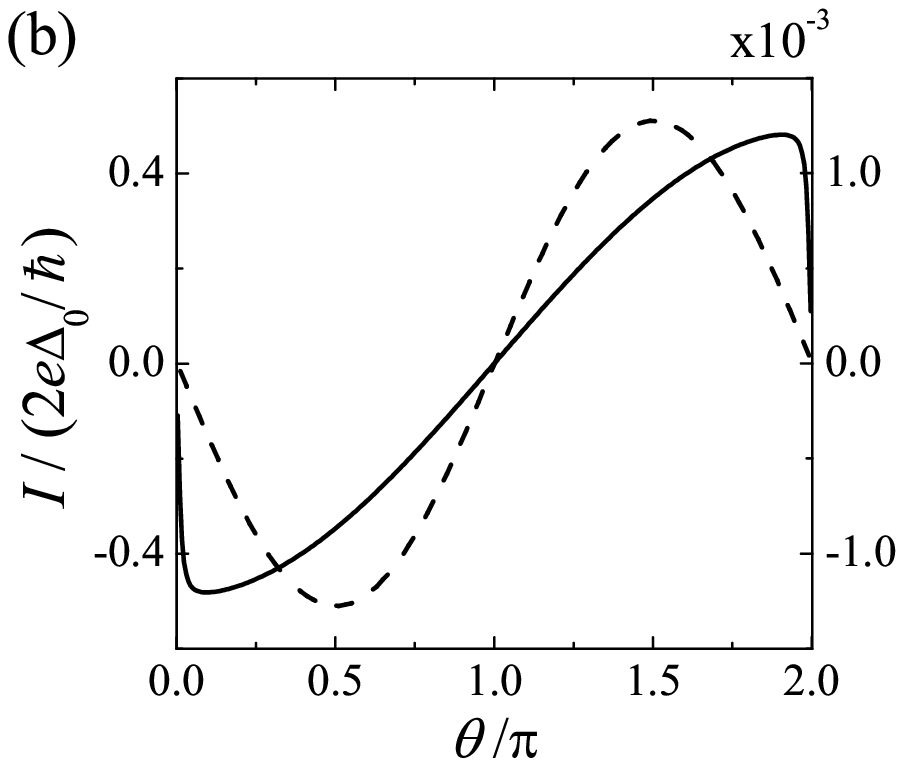}}
\caption{
(a) Free energy $F$ as a function of $\theta$, and (b) current-phase relation 
in a SC/FM/SC junction for $h_{ex} = 0.31 \mu_F$.  
The solid and the dashed lines are 
for the metallic contact ($Z=0$) and the tunnel junction ($Z=3$), respectively, 
and $\Delta_0$ is the superconducting gap at zero temperature.  
In these figures, the left vertical axes are for $Z=0$ and 
the right vertical axes are for $Z=3$.  
In addition, in (a) the origin of the vertical axes is arbitrary.}
\label{Fig_FE-I}
\end{figure}

Figures \ref{Fig_FE-I}(a) and \ref{Fig_FE-I}(b) show the $\theta$ dependence of the free 
energy $F$ and of the Josephson current $I=(2e/\hbar)({\partial F}/{\partial \theta})$ 
in the SC/FM/SC junction, respectively.  
Here, results are presented for two different values of the interfacial barrier 
$Z=mV/\hbar^2 k_F$, 
where $V$ is the strength of the $\delta$-function type of potential at the interfaces, 
$m$ is the electron mass, and $k_F$ is the Fermi wave number \cite{Furusaki}.  
The solid and the dashed curves in Figs. \ref{Fig_FE-I}(a) and \ref{Fig_FE-I}(b) 
correspond to the cases of $Z=0$ and $3$, respectively.  
In obtaining these results, we have assumed that 
the exchange field is $h_{ex} = 0.31 \mu_F$, where $\mu_F$ is the Fermi energy, 
the thickness of the FM is $L = 10/k_F$, 
and the coherence length at zero temperature is $\xi_0 = 1000/k_F$.  
As shown in Fig. \ref{Fig_FE-I}(a), $F$ has a minimum at $\theta = \pi$ ($\pi$ junction), and 
the variation of $F$ with $\theta$ is strongly dependent on $Z$.  
In the tunnel junction for $Z=3$, we have $F \approx - E_{\pi} \cos{(\theta+\pi)}$, 
where $E_{\pi}$ is the Josephson coupling constant.  
This gives the current with the $\pi$-shifted sinusoidal form 
which is shown in Fig. \ref{Fig_FE-I}(b) and 
described by $I \approx I_{\pi} \sin{( \theta + \pi )}$ with 
$I_{\pi}=2eE_{\pi}/\hbar$ the critical current.  
On the other hand, in the metallic contact ($Z=0$), 
we have approximately $F \approx - E_{\pi} \left| {\cos{((\theta + \pi)/2)}} \right|$.  
This leads to a current with the nonsinusoidal form 
$I \approx -(I_{\pi}/2)\sin{((\theta+\pi)/2)}$ for $0 \leq \theta \leq 2\pi$.  
The solid curve in Fig. \ref{Fig_FE-I}(b) shows that there are 
deviations from this form when $\theta$ is near $0$ or $2\pi$.  
However, these deviations do not affect the following discussion, 
and the approximate relations given here are valid for the cases of 
$0.28 \mu_F \lesssim h_{ex} \lesssim 0.34 \mu_F$ and 
$0.79 \mu_F \lesssim h_{ex} \lesssim 0.87 \mu_F$.  
In addition, it can be shown that these relations 
are valid for more realistic cases with small but finite $Z$ \cite{footnote}.  
Therefore, if we choose the appropriate values for $h_{ex}/\mu_F$ and $k_{F} L$, 
this form for the Josephson energy of the metallic $\pi$ junction 
is a good approximation, and in the following 
it will be used for the Josephson energy of the metallic $\pi$ junction 
in the superconducting ring.  

Now, we discuss the superconducting qubit which is shown in Fig. \ref{Fig_G}.  
In the ring, the SC/I/SC ($0$ junction) and the SC/FM/SC ($\pi$ junction) junctions 
have the Josephson energies 
$U_0  = - E_0 \cos{\theta_0}$ and 
$U_{\pi} = - E_\pi  \left| {\cos{((\theta_{\pi} + \pi)/2)}} \right|$, respectively.  
Here, $E_{0(\pi)}$ is the coupling constant in the $0$ ($\pi$) junction.  
The superconducting phase difference is 
$\theta_0$ for the $0$ junction, and $\theta_{\pi}$ for the $\pi$ junction.  
In this case, the total flux in the ring $\Phi$ satisfies the relation 
\begin{eqnarray}
\theta_{\pi} - \theta_0 = 2\pi \Phi/\Phi_0 - 2 \pi l, 
\end{eqnarray}
where $\Phi _0 = h/2e$ is the unit flux and $l$ is an integer.  
The total Hamiltonian of the ring is 
\begin{eqnarray}
H = K + U_0 + U_{\pi} + U_L, 
\end{eqnarray}
where $K = - 4E_{c} (\partial^{2}/\partial \theta_{0}^{2})$ is the electrostatic energy, 
$E_{c} = e^2/2C$ is the Coulomb energy for a single charge, 
$C$ is the capacitance of the $0$ junction, 
and $U_L$ is the magnetic energy stored in the ring.  
Here, the electrostatic energy in the $\pi$ junction is neglected, since 
the capacitance of the metallic $\pi$ junction 
is much larger than that of the insulating $0$ junction.  
The magnetic energy $U_L$ is given by 
$U_L  = {\left( {\Phi  - \Phi _{ext} } \right)^2 }/{2L _{s}}$, 
where $L _{s}$ is the self-inductance of the ring 
and $\Phi _{ext}$ is the external magnetic flux.  
The total Hamiltonian $H$ is analogous to 
that describing the motion of a particle with kinetic energy $K$ 
and in a potential $U_{tot} = U_0 + U_{\pi} + U_L$.  
Using the relation between the phase and the total flux, 
the potential $U_{tot}$ becomes a function of $\theta_{\pi}$ and $\Phi$: 
$U_{tot}=U_{tot}\left( {\theta_{\pi}, \Phi} \right)$.  
In order to obtain the state of the ring, 
we seek the solution at which $U_{tot}$ is minimum.
First, we minimize $U_{tot}$ with respect to $\Phi$, 
i.e., ${\partial U_{tot}}/{\partial \Phi} = 0$, which yields
\begin{equation}
\Phi \left( \theta_0 \right) = \alpha \frac{{\Phi_0}}{{2\pi}}\sin \theta_0 + \Phi _{ext}, 
\label{Phi-solution}
\end{equation}
where $\alpha = 2\pi {I_0 L _{s}}/{\Phi _0 }$ and 
$I_0 = 2eE_{0}/\hbar$ is the critical current in the $0$ junction.  
Substituting Eq. (\ref{Phi-solution}) in the expression for $U_{tot}$, we obtain
\begin{eqnarray}
U_{tot}/E_0 && = - \cos \theta_0 
                + \frac{\alpha}{2} {\sin^2{\theta_0} } \nonumber\\
     && - \beta \left| {\cos \left( {\frac{{\theta_0  + \pi }}{2} 
        + \frac{\alpha }{2}\sin \theta_0 
        + \pi \frac{{\Phi _{ext} }}{{\Phi _0 }}} \right)} \right| , 
\label{Utot}
\end{eqnarray}
where $\beta = E_{\pi}/E_0$.  

\begin{figure}[t]
\centering
{\includegraphics[width=0.85\columnwidth]{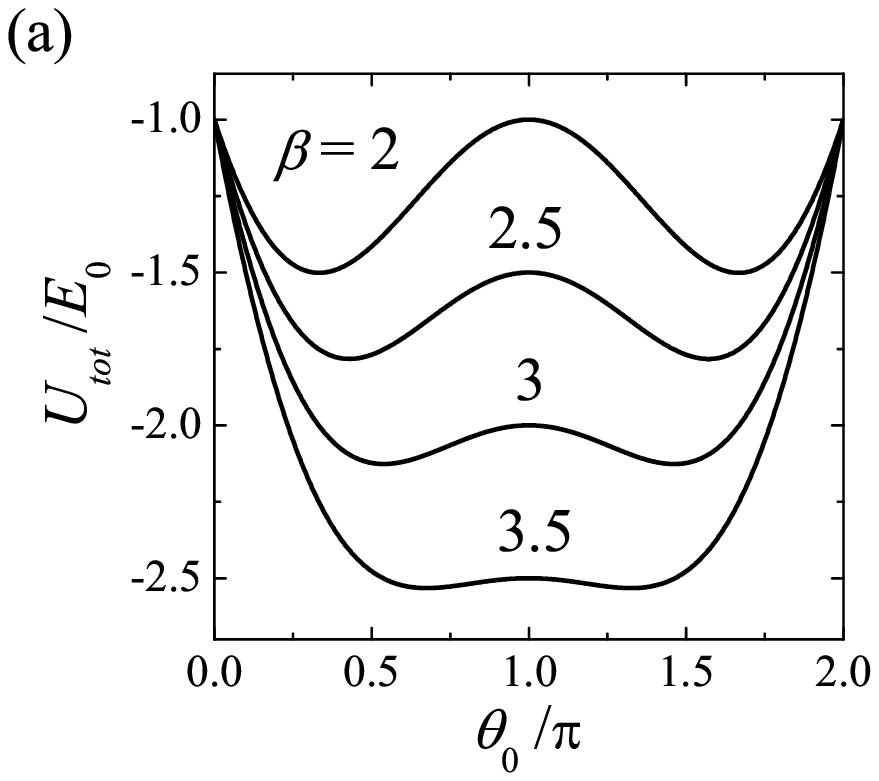}
\includegraphics[width=0.85\columnwidth]{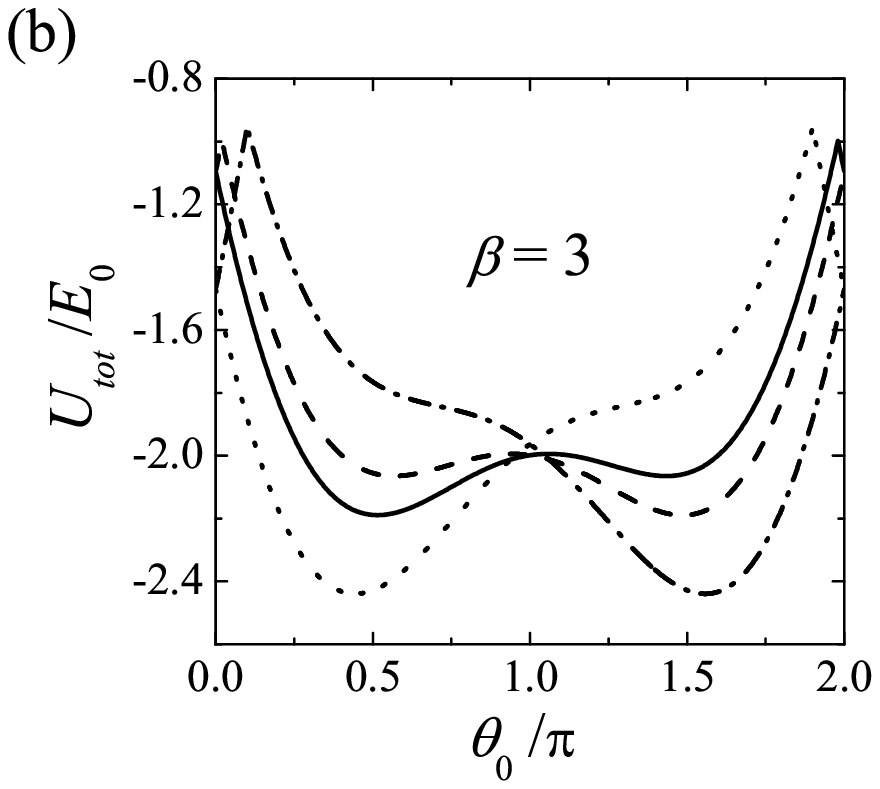}}
\caption{Normalized $U_{tot}$ as a function of $\theta_{0}$ 
for $\alpha = 3.1 \times 10^{-3}$ 
shown in (a) when there is no external magnetic field ($\Phi_{ext} = 0$), and 
shown in (b) when the field is finite: 
$\Phi_{ext}/\Phi_{0} = 0.01$ (solid line), $-0.01$ (dashed line), 
$0.05$ (dotted line), and $-0.05$ (dot-dashed line).}
\label{Fig_DM}
\end{figure}

Figure \ref{Fig_DM}(a) shows the $\theta_0$ dependence of the normalized $U_{tot}$ 
for various values of $\beta$ when there is no external magnetic field ($\Phi_{ext} = 0$).  
Here, we have taken $L _{s} = 2.0 \,{\rm pH}$ 
for the ring with a diameter of $2 \,{\rm \mu m}$ and $I_0 = 500 \,{\rm nA}$, 
which leads to $\alpha \approx 3.1 \times 10^{-3}$  \cite{Mooij,Chiorescu}.  
As shown in Fig. \ref{Fig_DM}(a), 
$U_{tot}$ has double minima located at $\theta_0 \approx \pi/2$ and $3\pi/2$, and 
the barrier height between the two degenerate states, 
$\left| \uparrow \right\rangle$ ($\theta_0 \approx \pi/2$) and 
$\left| \downarrow \right\rangle$ ($\theta_0 \approx 3\pi/2$), is controlled by $\beta$.  
The value of $\beta$ can be adjusted 
by changing the thickness and the area of the insulating barrier or of the ferromagnet.  
For $\left| \uparrow \right\rangle$ and $\left| \downarrow \right\rangle$ states, 
currents $I$ of magnitude $\lesssim I_0$ flow in the clockwise and counterclockwise directions, 
respectively, inducing flux $\Phi = L_{s}I \approx \pm 4.8 \times 10^{-4}\Phi_{0}$.  
Because of the quantum tunneling 
between $\left| \uparrow \right\rangle$ and $\left| \downarrow \right\rangle$, the bonding 
$\left| 0 \right\rangle \propto \left| \uparrow \right\rangle + \left| \downarrow \right\rangle$ 
and the antibonding 
$\left| 1 \right\rangle \propto \left| \uparrow \right\rangle - \left| \downarrow \right\rangle$ 
states are formed, hence yielding a two-level quantum system.  
For an ${\rm Al_{2}O_{3}}$ insulator with junction area $0.1 \,{\rm \mu m}^2$ 
and thickness $1 \,{\rm nm}$, 
it is formed that $E_{c} \approx 1.7 \times 10^{-24} {\rm J}$, and $E_{0}/E_{c} \approx 96$.  
In this case, from numerical calculations for $\beta = 3$, we estimate that 
the energy gap $\Delta E$ between the ground state $\left| 0 \right\rangle$ and 
the first exited state $\left| 1 \right\rangle$ 
is $\Delta E \approx 3.3 \times 10^{-24} {\rm J}$, which corresponds to 
a frequency of $\Delta E/h \approx 5.0 \,{\rm GHz}$.  
Microwave absorption measurements can be used to confirm this two-level quantum state.  
Figure \ref{Fig_DM}(b) shows the $\theta_0$ dependence of the normalized $U_{tot}$ 
for $\beta = 3$ within external magnetic fields.  
This figure shows that the degeneracy of the states 
$\left| \uparrow \right\rangle$ and $\left| \downarrow \right\rangle$ is lifted 
by applying a small external magnetic field to the ring.  
For $\Phi_{ext} = +(-) 0.01 \, \Phi_{0}$, 
the component of $\left| \uparrow \right\rangle$ ($\left| \downarrow \right\rangle$)
is larger than that of $\left| \downarrow \right\rangle$ ($\left| \uparrow \right\rangle$)
in the ground state $\left| 0 \right\rangle$, 
and vice versa in the first excited state $\left| 1 \right\rangle$.  
Within the presence of the larger external magnetic field ($\Phi_{ext} = \pm 0.05 \, \Phi_{0}$), 
the double-well potential disappears and the ground state is either 
$\left| \uparrow \right\rangle$ or $\left| \downarrow \right\rangle$.  
Therefore, finite currents flow in opposite directions 
for $\left| 0 \right\rangle$ and $\left| 1 \right\rangle$ states, 
when there is a finite external magnetic field.  
As a result of this, it is possible to distinguish the 
$\left| 0 \right\rangle$ and $\left| 1 \right\rangle$ states 
by measuring the current flowing in the ring 
with a superconducting quantum interference device (SQUID) placed around the ring.  
Our qubit has coherent states which require no external bias magnetic field, 
thus only a small external magnetic field is needed 
for the manipulation and the detection of the states, as compared to 
the half unit flux $\Phi_{0}/2$ required in the proposal of Ref. \cite{Mooij}.  
For instance, even if the dimension of the qubit is several 100 {\rm nm}'s, 
only magnetic fields of the order of a millitesla are needed for the manipulation of our qubit.  
This feature enables us to make qubits with smaller size which 
is advantageous in large-scale integration.  This type of qubit is also 
resistant to external noise and has longer decoherence time \cite{Mooij}.  
In order to realize the universal quantum logic gates, 
a controlled-NOT gate is needed in addition to the single qubit rotations discussed above.  
Using our qubit, the controlled-NOT gate is realized in the following way: 
When two qubits A and B have an inductive coupling, 
the energy gap in qubit A depends on the state of qubit B.  
In other words, the energy gap for qubit A is $\Delta E_{\rm A0}$ or $\Delta E_{\rm A1}$ 
when qubit B is in state $\left| 0 \right\rangle$ or $\left| 1 \right\rangle$, respectively.  
Now, if a microwave pulse with the frequency $\Delta E_{\rm A1}/h$ is applied to qubit A, 
the state of qubit A can be changed 
only if qubit B is in state $\left| 1 \right\rangle$.  
This is our proposal for the realization of the controlled-NOT gate.  

We also propose a qubit which consists of a superconducting ring with a metallic 
superconductor/normal metal/superconductor (SC/NM/SC) junction 
and a ferromagnetic $\pi$ junction.  
In the metallic SC/NM/SC junction, the phase dependence of the free energy 
is $\pi$-shifted from that of a metallic $\pi$ junction.  
The double minima are formed in the ring regardless of the interfacial barrier height in 
the $\pi$ junction, and hence the ring has potential as a qubit.  

The qubits which we propose have the following advantages: 
(i) the geometry is simple, i.e., only two Josephson junctions are used, 
(ii) the qubit is constructed without an external magnetic field, 
and only a small external magnetic field is needed in the measurement of the state.  
Furthermore, because of these advantages, 
(iii) the size of the qubit can be reduced.  
This makes large-scale integration possible, and 
leads to a reduction of the decoherence due to the coupling to the environment.  
Spin-electronic devices have been extensively studied.  
The qubits offer a new route for spin-electronics to quantum computing.  

\acknowledgements
We thank S.E. Barnes, N. Bulut, and H. Imamura for helpful and fruitful discussions.  
T.Y. was supported by JSPS Research Fellowships for Young Scientists.  
This work was supported by NAREGI Nanoscience Project, Ministry of 
Education, Culture, Sports, Science and Technology (MEXT) of Japan, 
and by a Grant-in-Aid from MEXT and NEDO of Japan.  



\begin{thebibliography}{99}
\bibitem{footnote} 
The value of $Z$ was estimated to be between $0$ and $\approx 0.31$ from measurements of 
the conductance at point contacts between FM and SC 
(see, for example, Ref. \cite{Strikers}).  
In our numerical calculations, we find that the form for the free energy $F$ 
given in the text is valid when $Z \lesssim 0.7$.
\end{thebibliography}
\end{document}